\documentclass{sig-alternate-05-2015}
\begin{document}

\setcopyright{acmcopyright}

\doi{http://dx.doi.org/xx.xxxx/xxxxxxx.xxxxxxx}

\isbn{978-1-4503-4486-9/17/04}


\acmPrice{\$15.00}

%
\conferenceinfo{SAC'17,}{ April 3-7, 2017, Marrakesh, Morocco}
\CopyrightYear{2017} 

\title{Towards Microservices and Beyond}
\subtitle{An incoming Paradigm Shift in Distributed Computing}
\numberofauthors{4}
\author{
\alignauthor
  Manuel Mazzara\\
         \affaddr{Innopolis University, Russia}\\
         \email{m.mazzara@innopolis.ru}
\alignauthor 
Ruslan Mustafin\\
         \affaddr{Innopolis University, Russia}\\
         \email{r.mustafin@innopolis.ru}
\and
\alignauthor 
Larisa Safina\\
         \affaddr{Innopolis University, Russia}\\
         \email{l.safina@innopolis.ru}
\alignauthor 
Ivan Lanese\\
         \affaddr{University of Bologna/INRIA}\\
         \email{ivan.lanese@gmail.com}
}
\date{30 July 1999}
\maketitle
\begin{abstract}
The microservice architecture is a style inspired by service-oriented computing that has recently started gaining popularity and that promises to change the way in which software is perceived, conceived and designed.  In this paper we offer a short overview intended as a collection of bibliographic references and links in the field of Microservices Science and Engineering (MSE).
\end{abstract}


%

%
%



\section{Background}
History of programming languages and paradigms have been characterized in the last few decades by a progressive shift towards distribution, modularization and loose coupling, with the purpose of increasing code reuse and robustness \cite{AlmeidaALGM04}, ultimately a necessity dictated by the need of increasing software quality, not only in safety and financial-critical applications, but also in more common off-the-shelf software packages. The two directions of modularization (code reuse and solid design) and robustness (software quality and formal methods: verification/correctness-by-construction) advanced to some extent independently and pushed by different communities, although with a non-empty overlap.

Object-oriented technologies are prominent in software development \cite{rumbaugh1991object}, with specific instances of languages incorporating both the aspects aforementioned (modularity and correctness). A notable example is the Eiffel programming language \cite{Meyer1988}, incorporating solid principles of OOP within a programming framework coordinated by the idea of design-by-contract, which aims at correctness-by-construction.  None of these technologies can nevertheless rule out the need for testing, which robustly remains a pillar of the software development lifecycle.  Other examples exist of languages having a strong emphasis on correctness, both from the architectural viewpoint and in terms of meeting functional requirements \cite{Mazzara2010}. However, until recently, not much attention was dedicated to integrating these principles into a distributed setting winning out properties such as easiness of deployment, a lightweight design and development phase, and minimal need for integration testing.

\section{Paradigm Shift}
Jolie~\cite{MGZ14} is a programming language functionally combining a multiplicity of aspects that are destined to revolution the way in which software is conceived, designed and understood. Originated from a major formalization effort~\cite{sensoria} for workflow and service composition~\cite{Mazzara11}, the language does not integrate a notion of correctness; it is simply built on it. The intuitiveness of the message-passing paradigm supports the design phase and avoids side effects that are not trivial to test.

As an open source project, Jolie has already built a community of developers worldwide - both in the industry and in academia - taking care of the development, continuously improving its usability, and therefore broadening the adoption. Recent developments are bringing the language to full maturity: extension of the type system \cite{Safina2016}, development of static type checking \cite{Tchitchigin2016}, addition of more iterative control structures to support programming, and inline automatic documentation \cite{Bandura2016} geared up the development environment and started the process of transforming it into a full suite that makes the entire concept attractive to developers and marketable to companies.

From the architectural point of view, Jolie has the potential to lead to a paradigm shift. Component-wise each building block is built as a microservice \cite{M16} embedding business capabilities in isolation. Every microservice can be reused, orchestrated, and aggregated with others~\cite{montesi}. This approach brings simplicity in components management, reducing development and maintenance costs, and supporting distributed deployments~\cite{fowler-tradeoffs}.

\section{Towards Microservices}

The shift towards microservices is a sensitive matter these days, seeing several companies involved in a major refactoring of their back-end systems to accommodate the easiness of the new paradigm. Other companies just start their business model developing software following the microservice paradigm since day one. We are in the middle of a major change in the view in which software is intended, and in the way in which capabilities are organized into components, and industrial systems are conceived.

The microservices architecture \cite{ms-book} is built on very simple principles:

\begin{itemize}
	\item \textit{Bounded Context}. First introduced in \cite{evans2004domain}, this concept captures one of the key properties of microservice architecture: focus on business capabilities. Related functionalities are combined into a single business capability which is then implemented as a service.
	\item \textit{Size}. Size is a crucial concept for microservices and brings major benefits in terms of service maintainability and extendability. Idiomatic use of microservices architecture suggests that if a service is too large, it should be refined into two or more services, thus preserving granularity and maintaining focus on providing only a single business capability.
	\item \textit{Independency}. This concepts encourages loose coupling and high cohesion by stating that each service in microservice architectures is operationally independent from others, and the only form of communication between services is through their published interfaces. 
\end{itemize}

\section{Microservices and Beyond}

The microservice architecture is a style that is increasingly  gaining popularity, both in academia and in the industrial world. Even though it is likely to conduct to a paradigm shift and a dramatic change in perception, it does not build on vacuum, and instead relates to well-established paradigms such as OO and SOA. In \cite{ms-book} a comprehensive survey on recent developments of microservice architecture is presented focusing on the \emph{evolutionary} aspects more than the \emph{revolutionary} ones. The presentation there is intended to help the reader in understanding the distinguishing characteristics of microservices.

Holding on the optimism the future is certainly not challenge-free. Security of the microservice paradigm is an issue almost fully untouched \cite{ms-book}. Commercial-level quality packages for development are still far to come, despite the acceleration in the interest regarding the matter. Fully-verified software is an open problem the same way it is for more traditional development models. That said, several research centers around the world have addressed and are addressing all these issues in the attempt to ride the wave and make the new generation of distributed systems a reality.

\small

\subsection*{Acknowledgements}
We would like to thank Innopolis University for logistic and financial support, and all the colleagues of the Service Science and Engineering and Software Engineering labs.


\normalsize



\bibliographystyle{abbrv}
\bibliography{bibliography}  
\end{document}